%% file: main.tex
\documentclass[letterpaper, 10 pt, conference]{ieeeconf}  
\IEEEoverridecommandlockouts                              
\usepackage{microtype}

\usepackage{graphics} 
\usepackage{epsfig} 
\usepackage{amsmath} 
\usepackage[sort,compress]{cite}%

\usepackage[pdftex,pdftitle={Machine Learning-based Estimation of Forest Carbon Stocks to increase Transparency of Forest Preservation Efforts}]{hyperref}
\hypersetup{colorlinks,linkcolor={green!50!black},citecolor={green!50!black},urlcolor={black}, pdfauthor={Bj{\"o}rn L{\"u}tjens, Lucas Liebenwein, Katharina Kramer}}
\makeatletter \let\NAT@parse\undefined \makeatother
\usepackage[sort,compress]{cite}
\usepackage{graphicx} 
\usepackage{amsfonts}
\usepackage{amsmath,soul}
\usepackage{color}
\usepackage[font=small]{subcaption}
\usepackage{balance}
\usepackage[font=small]{caption}
\usepackage[linesnumbered,ruled,vlined]{algorithm2e}
\usepackage{multirow}
\usepackage{dsfont}

\usepackage{tabulary}
\newcolumntype{K}[1]{>{\centering\arraybackslash}p{#1}}

\iffalse
\newif\ifPDF \ifx\pdfoutput\undefined\PDFfalse \else\ifnum\pdfoutput > 0\PDFtrue \else\PDFfalse \fi \fi

\ifPDF
\usepackage[pdftex, pdfstartview={FitV}, pdfpagelayout={TwoColumnLeft},bookmarksopen=true,plainpages = false, colorlinks=true, linkcolor=black, citecolor = black, urlcolor = black,filecolor=black, pagebackref=false,hypertexnames=false, plainpages=false, pdfpagelabels ]{hyperref}
\fi

\usepackage[capitalize]{cleveref}
\crefformat{equation}{(#2#1#3)}%
\Crefformat{equation}{Equations~(#2#1#3)}
\Crefname{equation}{Equation}{}

\makeatletter
\def\bstctlcite#1{\@bsphack
  \@for\@citeb:=#1\do{%
    \edef\@citeb{\expandafter\@firstofone\@citeb}%
    \if@filesw\immediate\write\@auxout{\string\citation{\@citeb}}\fi}%
  \@esphack}
\makeatother

\title{\LARGE \bf Machine Learning-based Estimation of Forest Carbon Stocks to increase Transparency of Forest Preservation Efforts}

\author{Bj{\"o}rn L{\"u}tjens, Lucas Liebenwein and Katharina Kramer\\
Massachusetts Institute of Technology
}

\usepackage[svgnames]{xcolor} \definecolor{DarkGreen}{rgb}{0,0.5,0}
\definecolor{DarkRed}{rgb}{0.75,0,0}

\usepackage[final]{changes} 
\definechangesauthor[name={J.~H.}, color={blue}]{jh}
\definechangesauthor[name={M.~E.}, color={red}]{me}
\definechangesauthor[name={B.~L.}, color={green}]{bl}
\definechangesauthor[name={G.~H.}, color={orange}]{gh}

\usepackage{tikz,mathtools}
\usepackage[capitalize]{cleveref}
\crefformat{equation}{(#2#1#3)}
\Crefformat{equation}{Equation~(#2#1#3)}
\Crefname{equation}{Equation}{Equations}

\usetikzlibrary{shapes,positioning,automata,arrows,fit,backgrounds,calc}
\tikzstyle{block} = [draw, fill=blue!20, rectangle,minimum height=1em,
minimum width=2em] \tikzstyle{sum} = [draw, fill=blue!20, circle, node
distance=1cm] \tikzstyle{input} = [coordinate] \tikzstyle{output} =
[coordinate] \tikzstyle{pinstyle} = [pin edge={to-,thin,black}]
\usetikzlibrary{trees} \usetikzlibrary{decorations.pathmorphing}
\usetikzlibrary{decorations.markings}
\definecolor{darkgreen}{rgb}{0,0.5,0}
\definecolor{darkred}{rgb}{220,20,60}

\begin{document}


\maketitle
\thispagestyle{empty}
\pagestyle{empty}

\begin{abstract}
An increasing amount of companies and cities plan to become CO2-neutral, which requires them to invest in renewable energies and carbon emission offsetting solutions. One of the cheapest carbon offsetting solutions is preventing deforestation in developing nations, a major contributor in global greenhouse gas emissions. However, forest preservation projects historically display an issue of trust and transparency, which drives companies to invest in transparent, but expensive air carbon capture facilities. Preservation projects could conduct accurate forest inventories (tree diameter, species, height etc.) to transparently estimate the biomass and amount of stored carbon. However, current rainforest inventories are too inaccurate, because they are often based on a few expensive ground-based samples and/or low-resolution satellite imagery. LiDAR-based solutions, used in US forests, are accurate, but cost-prohibitive, and hardly-accessible in the Amazon rainforest. We propose accurate and cheap forest inventory analyses through Deep Learning-based processing of drone imagery. The more transparent estimation of stored carbon will create higher transparency towards clients and thereby increase trust and investment into forest preservation projects.  
\end{abstract}

\input{intro}







\clearpage
\section{Acknowledgements} \label{sec:ack}
The authors want to thank La Niebla Forest for hospitality and support in the local community; World Wildlife Fund (WWF) Peru, Peru Ministry of Agriculture - National Wildlife and Forest Service (SERFOR), Peru Ministry of Environment - National Forest Conservation Program (BOSQUES), Peru National Geographics Institute (IGN), VividEconomics, WeRobotics, and UAV Peru for helpful discussions about the difficulties of reforestation and forest conservation; Prof. Newman, Prof. Wood, Prof. Fernandez, Prof. How, and Prof. Rus for their advice on remote sensing, UN politics, carbon sequestration, and robotics; MIT Sandbox Innovation Fund, MIT PKG IDEAS Global Challenge, MIT Legatum Seed Travel Grant, Microsoft AI for Earth Grant, and NASA Space Grant for their support. The work is conducted by the Sustainable AI Initiative {\tt\{sustainable-ai.mit.edu\}}, at the Massachusetts Institute of Technology, 77 Mass.\ Ave., Cambridge, MA, USA. {\tt lutjens@mit.edu}.

\balance

\bibliographystyle{IEEEtran} 
{\tiny
\bibliography{biblio}
}

\end{document}

%% file: intro.tex

\section{The Problem} \label{sec:problem}
Deforestation and forest degradation are responsible for ${\sim}15\%$ of global greenhouse gas emissions, as burning forest releases stored carbon into the air~\cite{FCPF_2018,UN_redd_2018,IPCC_2014}. 
Stopping deforestation and forest degradation and supporting sustainable forestry hence mitigates climate change and also preserves biodiversity, prevents flooding, controls soil erosion, reduces river siltation, and offers a workplace for the local population~\cite{FCPF_2018}. 
Despite the paramount importance of reforestation and preservation efforts, they are far from sufficient, mostly because of a lack of financing~\cite{Interviews_2018,FCPF_2018}. 
This financial gap is created by a severe lack of trust into reforestation and preservation projects as they are not transparent in their CO2 impact to client companies that want to offset carbon emissions~\cite{Alcoa_2017,Interviews_2018}. 
\begin{figure}[t]
	\centering
    \includegraphics [trim=0 0 0 0, clip, width=1.0\columnwidth, angle = 0]{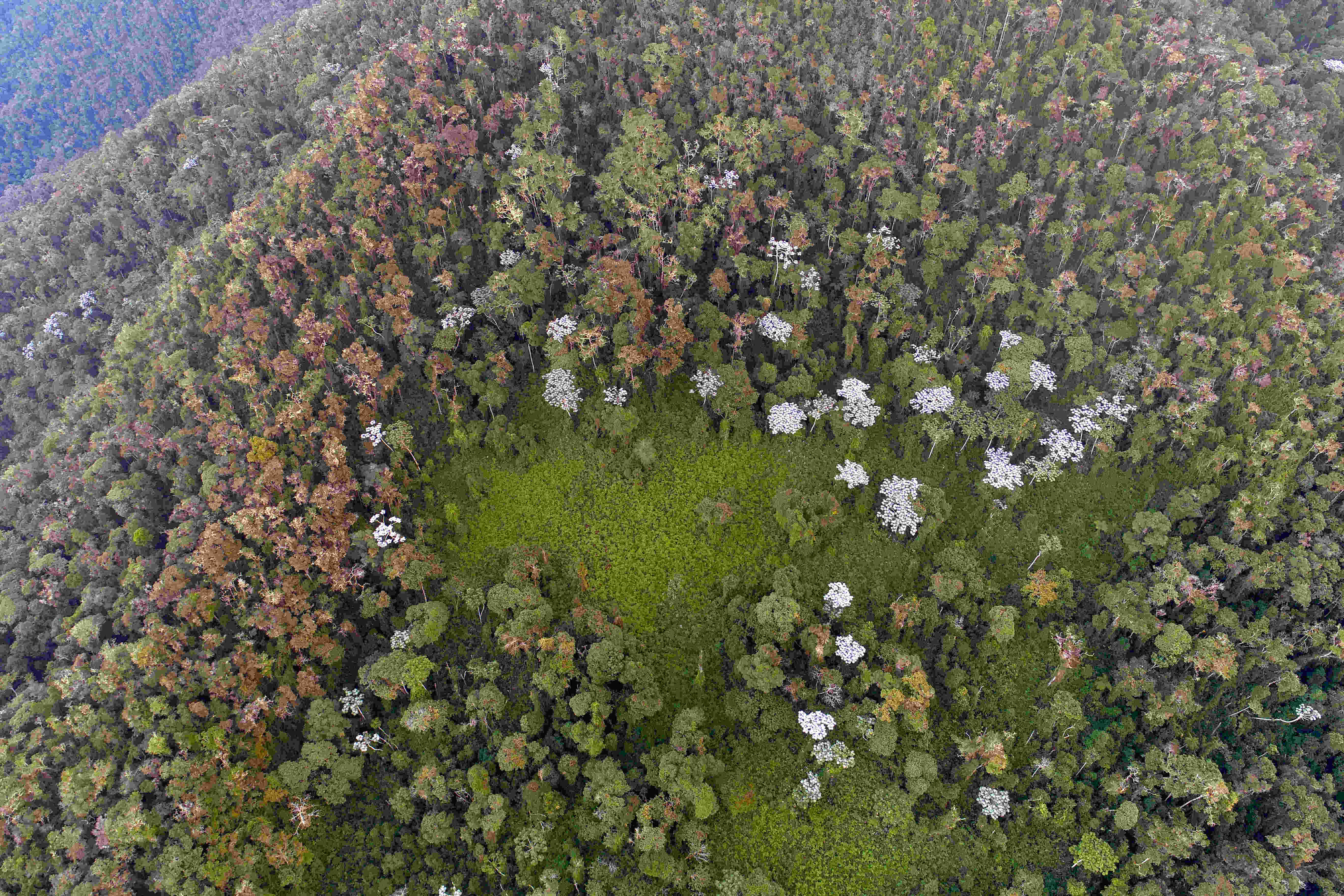}
    \label{fig:example_sat} 
	\caption{Medium-resolution drone imagery, collected during pilot flights near San Ramón, Perú.}
\vspace{-0.2in}
\end{figure}

Forest inventories are common practice in forestry, account for $25\%$ of the expenses of reforestation and estimate stored carbon~\cite{Interviews_2018}. Apart from carbon estimation, forest inventories are also created to identify illegal logging, control pests and diseases, estimate the opportunity cost of deforestation, manage wildfire hazards, and achieve sustainable forestry.
Classical forest inventories are created through manually counting and classifying trees in a ${\sim} 7{-}15$m radius every $0.09{-}1$km~\cite{Interviews_2018}. The sparse samples are interpolated, recently with the help of satellite imagery
, to create an inventory for the whole forest. Ground-based sampling, however, is prohibitively expensive (${\sim}300$USD/ha), and time-intensive ($2{-}7$ days/$20$ha) in large-scale rainforests, due to dense vegetation, a large team of experts, and scarcity of roads~\cite{Interviews_2018}. 

Purely satellite-based approaches mostly use publicly available RGB-NIR satellite imagery, or radar. As the low-resolution (RGB max. $30$cm/px, radar ${\sim}250$m/px) does not suffice to accurately determine the tree count, species, or height, most satellite-based approaches only measure area covered by forest which leads to rough estimates of carbon sequestering potential with high uncertainties~\cite{Gibbs_2007,Interviews_2018}. 
LiDAR-based approaches, used in US forests, are very accurate, but hardly-accessible and cost-prohibitive for low budget reforestation and preservation projects, because of the expense of the sensor and the bigger carrying drone, or plane~\cite{Zolkos_2013, Interviews_2018}.


\section{The Solution/Innovation} \label{sec:solution}
Our goal is to increase investment into reforestation and preservation projects to combat climate change by providing an accurate, cheap, and transparent carbon storage analysis. The analysis is supplied to reforestation and preservation projects that, with the analysis, have sufficient trust to convince their client companies to higher investments.
\subsection{Technical Solution} \label{sec:tech_sol}
\begin{figure}[t]
	\centering
    \includegraphics [trim=0 0 0 0, clip, width=1.0\columnwidth, angle = 0]{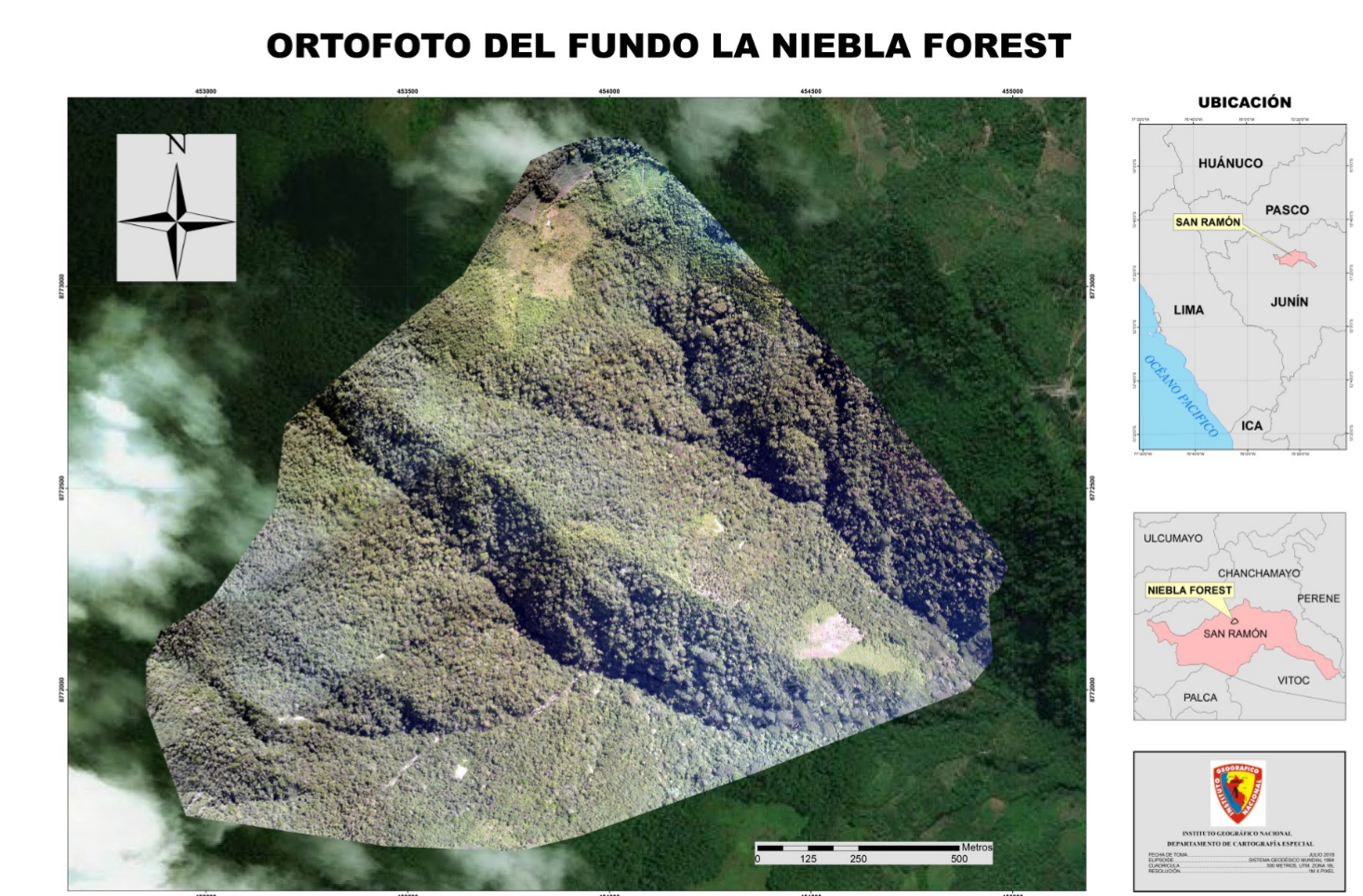}
  \caption{Collected map from pilot flights with the National Geographics Institute of Peru near San Ramón.}
  \label{fig:map} 
\vspace{-0.25in}
\end{figure}

The proposed forest inventory assessment consists of an on-site data collection and an off-site processing part. During the data collection with the local partner, a low-cost quadrotor (DJI Phantom 4 Pro, $1.5$kUSD) and five batteries have to been used to map $100$ha in ${\sim}5$hrs with one operator (${\sim}10$USD/ha) for~\cref{fig:map}. DroneDeploy was used to plan the flight and mosaic the images. The next iteration will be an off-the-shelf, low-cost vertical take off and landing (VTOL) fixed-wing drone to cover up to $250$ha in one $60$min flight and launch in dense forests. The drone will be equipped with a gimbal, 4k GoPro RGB camera, and a Sentera NDVI-IR camera. 

Deep Learning algorithms are proposed to extract crown diameter, species, and count of emergent and canopy trees. Specifically, a pixel-wise segmentation algorithm, based on DeepLabv3+~\cite{Chen_2018}, a Convolutional Neural Network (CNN) architecture, will classify the tree species at each pixel of the collected RGB-NIR imagery and extract crown diameter and tree count. The expected success of the algorithm assumes that a canopy's RGB-NIR spectrum and shape strongly correlate with the tree species. The correlation is shown for high-resolution sensors in~\cite{Cochrane_2000,Vauhkonen_2009}, but needs to be validated with the available low-cost sensors in future results. Additionally, a Bayesian regression model with spatial random effects~\cite{Finley_2007} with the same in- and outputs is being developed to increase overall accuracy via model ensembling, and counteract the inaccuracy of the CNN model on novel data. 

In addition to crown diameter and species, the estimation of forest carbon stocks requires canopy heights (distance from ground to canopy). Canopy heights cannot be accurately inferred from drone imagery, because visibility of the forest floor is prohibited by dense vegetative cover. Hence, a digital surface model (DSM) of the surface heights (distance from sea level to canopy), based on GPS, IMU, and structure from motion was created with the DroneDeploy software. A satellite-based digital elevation/terrain model (DEM) (distance from sea level to ground) will be subtracted from the DSM to obtain the canopy height model (CHM). The accuracy of the approach will be benchmarked on ground-based inventories. 

Allometric equations can be used to calculate forest biomass and carbon stocks, from canopy height, crown diameter, and species~\cite{Gold_standard,Gibbs_2007}. The accuracy of multiple allometric equations for tropical rainforest, and Andean rainforests that do or do not contain information about the tree species~\cite{Jerome_2014} will be evaluated. 

An accurate, but small dataset~\cite{ECODSE_2017} with tree height, species and crown segmentation is used. A larger dataset will be created by fusing ground-based and remotely sensed inventories of well studied forests (e.g., US national forests~\cite{NSF_2018}). 
\subsection{Partnerships} \label{sec:location}
\begin{itemize}
  \item A very close connection to a local community partner, which offers 100 hectares of rainforest in San Ramón, Perú as testing ground has been established. The community partner visits local mayors, and schools, and creates social media initiatives to reduce deforestation. The partner has started a small-scale reforestation project.
  \item NGOs and ministries have been visited to access data, co-develop software, and deploy it at scale
  \item We are continuously reaching out to gain knowledge in Forestry, Citizen Science, and Remote Sensing.
\end{itemize}

\subsection{Scalability} \label{sec:scalability}

As the approach is scaled to larger areas of forest, the local communities will be involved in the monitoring of preservation projects to make them feel responsible and technologically capable to protect their forest. To do so, an app will be developed that allows locals to map forests and scale up the data collection nationally. The app will be rolled out to the community partners' network of volunteers and local municipalities that possess a drone.

For the long-term, the cheap, and accurate ML-based carbon inventories are proposed to be embedded as standard in the cap-and-trade carbon market. The California Air Resources Board currently considers a bill to integrate CO2 offsets from tropical reforestation. This would allow reforestation and preservation projects to earn $10{-}15$ USD per ton of sequestered CO2 and incentivize locals, strongly concerned about monetary aspects, to sustain primary forests. Forests would be a competitive carbon offsetting choice, because they store a ton of CO2 at roughly $20{-}25$USD ($6{-}8$ trees; one tree costs ${\sim}3$ USD ($30\%$ seedling, $45\%$ labour, $25\%$ monitoring))~\cite{Interviews_2018}, whereas carbon capturing plants convert CO2 at a price of $94{-}232$ USD/tCO2~\cite{Keith_2018}.

The proposed method to infer forest inventories can also help reduce illegal logging. Timber companies are alloted internationally salable trees based on forest inventories of their land. The inventories, however, can be untruthfully overestimated, and companies sell rare and valuable trees from outside of their land. The proposed method can be used to cheaply verify the reported inventories of tree species.
\section{Impact}
Although mitigating climate change is this project's main goal, success is measured via the UN sustainable development goal 15.1.1, the ``ratio of total land covered by forest``, to incorporate the beneficial side effects of forest cover.
As this project is trying to increase the amount of trust und understanding that people have for carbon offsetting initiatives, e.g. reforestation, it is trying to change the bigger system. While at the beginning, it would be a success to increase investment into one offsetting project, the project aims for a large scale impact where people are more aware of how much effort it takes to offset their emissions, make them more environmentally conscious, and make investments into reforestation for carbon offsetting a standard. 
\subsection{Ethical considerations} \label{sec:ethical_consideration} 
\begin{itemize}
  \item An accurate forest inventory must be stored securely to prevent misuse for finding and logging rare trees
  \item Best practices for wildlife monitoring are respected~\cite{Hodgson_2016}
  \item Drone flights must be restricted via GPS to only fly over approved government or private land   
\end{itemize}